# Double well heterostructures for electric modulation of light

B. Laikhtman, S. Suchalkin and G. Belenky

The system of double quantum wells separated by barriers is suggested for switching and modulation of light. The system has potential for high operational speed and large modulation depth.

**Introduction**

Modulators and switches are the main components of the optical and THz communication technology. The high speed data communication needs switches and modulators operating with significant modulation depth and high frequencies. Over the past two decades many schemes based on different principles were suggested to meet these requirements. [1-4] Modulation depth in some devices is close to and even reaches 100%. [5-8]. The main mechanism of modulation is typically laser pumping of free carriers that changes transmittance and reflectivity of the material. The speed of modulation is limited by the carrier lifetime of the substrate so that modulation speed typically is of the order of kHz and in some devices reaches 200 kHz.[9] Another mechanism, modification of plasma frequency of localized surface plasmons with help of high energy pump laser, makes it possible to increase modulation speed up to 1.2 GHz at the expense of lower change of the transmittance and hence the modulation depth. [10]

In this paper we consider a structure that can be used for modulation and switching of optical or THz signal and combines high modulation depth with high modulation speed. The structure is 1D periodic systems of double quantum wells separated by impenetrable barriers. The optical absorption threshold is the energy gap between the highest heavy hole level in one well and the lowest electron level in the other. The system is a superlattice that differs from the regular ones by the barriers that prevent formation of 3D spectrum from 2D spectra of each period. The existence of the barriers makes it possible to apply electric field in the growth direction without current and accompanying high field instabilities. Such a field changes the energy difference between heavy hole levels in one well and electron levels in the other [11] and in this way it changes the threshold of optical absorption. 2D density of states in each period does not depend on energy. As a result, when the absorption threshold for given optical energy is crossed, the absorption coefficient does not grow gradually with the electric field but rises sharply from a negligible to a finite value providing large modulation depth. The structure can be used for electric modulation of both amplitude and phase [12] of the signal.

The modulation of signal with electric field in the proposed structure does not require change of carrier concentration. The speed of modulation is limited only by speed of application of electric field. So, high operational speed comparable that of Pockels and Kerr electrooptical effects [13-17] can be realized.

Essentially, the proposed structure is a modification of single well structures where the quantum confined Stark effect was demonstrated. [18-22] The advantage of double well compared to single well structures are significantly larger Stark effect that can be controlled by design and more flexibility in choice of optical parameters such as the absorption threshold.

The range of the electric field where the absorption coefficient changes from a negligible to a relatively large value depends on the electron scattering time. For relaxation time $10^{-11}$s the energy width of the transition interval is around 0.35 meV. If the period of the structure ~ 10 nm the field necessary to cross the transition ~ 0.7 kV/cm. Depending on the material and the width of the layers the considered structure can be used for modulation in a wide range of optical frequencies from mid IR down to THz.

The sharp change of the absorption coefficient in crossing of the absorption threshold with help of electric field is accompanied with peak of the refractive index. The variation of the refractive index in the peak is much larger than due to electrooptical effect in comparable field in such materials as $LiNbO_3$.

**Structure**

The basis of the whole consideration is the possibility to control the electron – heavy hole band gap in a double quantum well, Fig. 1, with external electric field. An external electric field *F* applied in the growth direction changes the energy band gap between the electron level in left well $\varepsilon_e$ and heavy hole level at the right well $\varepsilon_{hh}$ by, roughly speaking, the field multiplied by the distance between the middles of the wells, i.e., $eF(d_1 + d_2)/2$ where $d_1$ and $d_2$ are the widths of the left and right wells respectively.[11] E.g., if the distance between the centers of the wells is 5 nm then the change of the field from -50 kV/cm to +50 kV/cm changes the separation between heavy hole level at the right well and conduction band at the left well by about 50 meV. Thus if the difference between the band gap and the light quantum energy $|\varepsilon_e - \varepsilon_{hh} - \hbar\omega|$ is smaller than 25 meV then variation of the field makes it possible to cross the light absorption threshold.

The density of state in 2D case does not depend on the energy. Therefore above the threshold the absorption coefficient does not grow from zero with $\hbar\omega - (\varepsilon_e - \varepsilon_{hh})$ as this would be in 3D case but immediately reaches a finite value. The width of the threshold depends on the energy uncertainty due to carrier scattering that controls smearing of band edges.

For calculation of the absorption coefficient and refractive index we use two-band Hamiltonian of the heterostructure where *z* axis is along the growth direction, the right and left wells occupy regions $-d_1 < z < 0$ and $0 < z < d_2$ respectively.[23,24] The Hamiltonian includes just the conduction and heavy hole bands because the effect of light holes at the vicinity of the threshold is negligible.

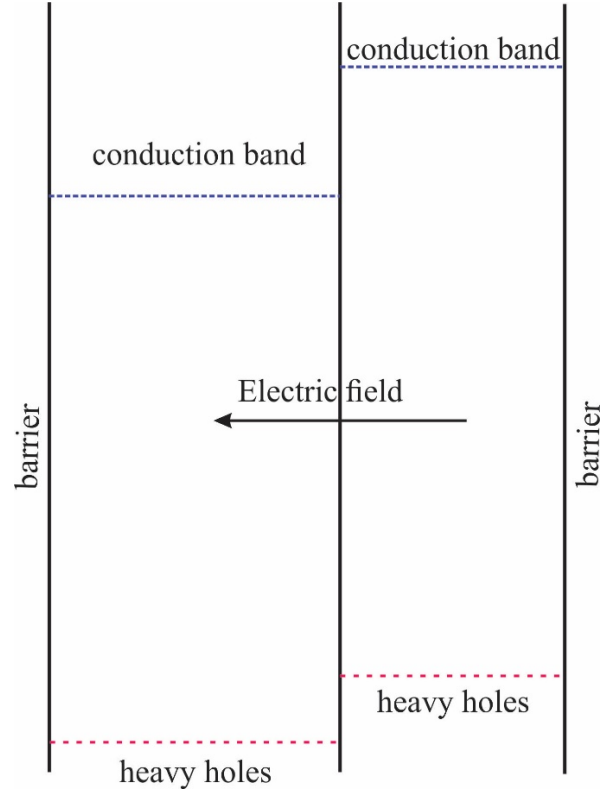

Fig. 1. Double quantum well structure in electric field.

$$H = \begin{pmatrix} \varepsilon_e & \dfrac{\hbar k_+ P}{\sqrt{2}m_0} \\ \dfrac{\hbar k_- P^*}{\sqrt{2}m_0} & \varepsilon_{hh} \end{pmatrix} \qquad (1)$$

Here $k_\pm = k_x \pm i k_y$ are components of in-plane wave vector, $m_0$ is the free electron mass,

$$P = P_1 \int_{-d_1}^{0} \xi_e^*(z)\xi_{hh}(z)dz + P_2 \int_{0}^{d_2} \xi_e^*(z)\xi_{hh}(z)dz \ , \qquad (2)$$

$P_1$ and $P_2$ are momentum matrix elements between the conduction and valence bands in the left and right wells, $\xi_e(z)$ and $\xi_{hh}(z)$ are envelope wave functions of electron and heavy hole levels in the double well systems at $k=0$.

The spectrum of Hamiltonian (1) is

$$E_{\pm} = \frac{\varepsilon_e + \varepsilon_{hh}}{2} \pm \sqrt{\frac{(\varepsilon_e - \varepsilon_{hh})^2}{4} + \frac{\hbar^2 |P|^2 k^2}{2m_0^2}} \, . \tag{3}$$

Its eigenfunctions are

$$\Psi_{\pm} = \begin{pmatrix} \psi_{e\pm} \\ \psi_{hh\pm} \end{pmatrix}, \qquad \psi_{e\pm} = |\psi_{\pm}| e^{i\varphi_{e\pm}}, \qquad \psi_{hh\pm} = |\psi_{\mp}| e^{i\varphi_{hh\pm}}, \tag{4}$$

where

$$\psi_{\pm} = \left[ \frac{\sqrt{\frac{(\varepsilon_e - \varepsilon_{hh})^2}{4} + \frac{\hbar^2 |P|^2 k^2}{2m_0^2}} \pm \frac{\varepsilon_e - \varepsilon_{hh}}{2}}{2\sqrt{\frac{(\varepsilon_e - \varepsilon_{hh})^2}{4} + \frac{\hbar^2 |P|^2 k^2}{2m_0^2}}} \right]^{1/2}, \tag{5}$$

and

$$\varphi_{e+} - \varphi_{hh+} = \varphi_k \, , \qquad \varphi_{e-} - \varphi_{hh-} = \varphi_k \pm \pi \, , \qquad \tan \varphi_k = \frac{k_y}{k_x} \, . \tag{6}$$

Whatever are optical properties of a double quantum well, the effect of single double well structure is quite weak. The structure that we are considering is a system of double quantum wells, i.e., a superlattice where each period is a double quantum well separated from the next one by a barrier. The barrier is necessary to prevent a current under electric field in the growth direction, so that each double well works independently of all others. On the other hand, for light beam along the growth direction the effects of periods are accumulated.

**Optical properties**

This system under consideration is characterized by a dielectric constant in the growth direction,

$$\kappa(\omega) = \kappa_{\infty} + 4\pi\chi(\omega) \tag{7}$$

where $\chi(\omega)$ is a susceptibility due to transitions between heavy hole band in the right well and conduction band in the left well of all periods of the superlattice and $\kappa_{\infty}$ is the dielectric constant resulting from transitions between all other bands. Absorption coefficient is connected with $\chi(\omega)$ by the relation

$$\alpha(\omega) = \frac{4\pi\omega}{cn} \operatorname{Im} \chi(\omega), \tag{8}$$

where $n = \sqrt{\kappa_\infty + 4\pi \operatorname{Re} \chi(\omega)}$ is the refractive index. The simplest way of calculation of the absorption coefficient and refractive index is to calculate $\alpha(\omega)$ in the Born approximation and then to recover $\operatorname{Re} \chi(\omega)$. The interaction Hamiltonian that provides transitions between conduction and valence band is $H_{\text{int}} = -(e/m_0 c)\mathbf{p}\mathbf{A}$ where $\mathbf{p}$ is the electron momentum operator and $\mathbf{A}$ is the vector potential of the light wave. As a result,

$$\alpha(\omega) = \frac{4\pi e^2}{dm_0^2 cn\omega} \sum_s \int |M_s(k)|^2 \frac{\hbar/\tau}{\left[E_+(k) - E_-(k) - \hbar\omega\right]^2 + \hbar^2/\tau^2} \frac{d^2k}{(2\pi)^2} \qquad (9)$$

where $d = d_1 + d_2 + d_b$ is the period of the superlattice, $d_b$ is the width of the barrier, the summation is taken over electron spin states, $M_s(k)$ is the matrix element of $\mathbf{ep}$ between initial and final electron states and $\mathbf{e}$ is the polarization vector. $\delta$ function in Eq.(9) is replaced with Lorentz function with carrier relaxation time $\tau$ that is important near the absorption threshold. Then the susceptibility is

$$\chi(\omega) = \frac{e^2}{dm_0^2 \omega^2} \sum_s \int |M_s(k)|^2 \frac{1}{E_+(k) - E_-(k) - \hbar\omega - i\hbar/\tau} \frac{d^2k}{(2\pi)^2} \qquad (10)$$

Complete wave functions necessary for calculation of the matrix element are

$$\Psi_{\mathbf{k}s-}(\mathbf{r}) = \frac{1}{\sqrt{S}} e^{i\mathbf{k}\mathbf{r}_\parallel} \left[ \psi_{hh-}(k)\xi_{hh}(z)u_{hh,s}(\mathbf{r}) + \psi_{e-}(k)\xi_e(z)u_{c,s}(\mathbf{r}) \right], \qquad (11a)$$

$$\Psi_{\mathbf{k}s+}(\mathbf{r}) = \frac{1}{\sqrt{S}} e^{i\mathbf{k}\mathbf{r}_\parallel} \left[ \psi_{hh+}(k)\xi_{hh}(z)u_{hh,s}(\mathbf{r}) + \psi_{e+}(k)\xi_e(z)u_{c,s}(\mathbf{r}) \right]. \qquad (11b)$$

Here $\mathbf{r}_\parallel$ is the in-plane radius vector, $S$ is the normalization area, $u_{hh,s}(\mathbf{r})$ and $u_{c,s}(\mathbf{r})$ are Bloch amplitudes in heavy hole and conduction band. Direct calculation gives

$$M_{1/2}(k) = P\left[ \frac{e_x - ie_y}{\sqrt{2}} \psi_{hh+}^*(k)\psi_{e-}(k) + \frac{e_x + ie_y}{\sqrt{2}} \psi_{e+}^*(k)\psi_{hh-}(k) \right], \qquad (12a)$$

$$M_{-1/2}(k) = -P\left[ \frac{e_x + ie_y}{\sqrt{2}} \psi_{hh+}^*(k)\psi_{e-}(k) + \frac{e_x - ie_y}{\sqrt{2}} \psi_{e+}^*(k)\psi_{hh-}(k) \right]. \qquad (12b)$$

The integral in Eq.(10) diverges at large $k$, i.e., the main contribution to this part comes from the region where $k$ is of the order of the inverse lattice constant $a_0$ and $\mathbf{kp}$ method that is the basis of the whole calculation is not valid. On the other hand, the divergent part does not depend on $\omega$ and can be considered as a part of $\kappa_\infty$. The final results are

$$\operatorname{Re}\chi(\omega) = \frac{e^2}{8\pi\hbar\omega d}\left[1+\frac{(\varepsilon_e - \varepsilon_{hh})^2}{(\hbar\omega)^2}\right]\ln\frac{|\varepsilon_e - \varepsilon_{hh}|}{\sqrt{(|\varepsilon_e - \varepsilon_{hh}|-\hbar\omega)^2 + \hbar^2/\tau^2}}, \qquad (13)$$

$$\alpha(\omega) = \frac{e^2}{2cn\hbar d}\left[1+\frac{(\varepsilon_e - \varepsilon_{hh})^2}{(\hbar\omega)^2}\right]\arctan\frac{\hbar/\tau}{|\varepsilon_e - \varepsilon_{hh}|-\hbar\omega}. \qquad (14)$$

It is noteworthy that the susceptibility does not depend on the momentum matrix element $P$. This happens because this matrix element enters both the transition matrix element $M_s(k)$ and the spectrum. That is an increase of $P$ leads to an increase of $M_s(k)$ but also to a decrease of the substantial part of the integration interval that is controlled by the spectrum. These two effects cancel each other. As a result, this part of the integral does not depend on $P$.

For this reason the maximal value of one period absorption,

$$\alpha_{dw,\max} = \frac{\pi e^2}{c\hbar n} = \frac{0.0229}{n}, \qquad (15)$$

that is reached near absorption threshold, $\hbar/\tau \ll \hbar\omega - |\varepsilon_e - \varepsilon_{hh}| \ll \hbar\omega$, depends on only one material parameter, the average refractive index.

The region of $k$ that contributes to the integral is defined by the relation $\hbar Pk/m_0 \sim |\varepsilon_e - \varepsilon_{hh}|$ Inasmuch **kp** method is valid only at $ka_0 \ll 1$ this applies a limitation of the validity of Eqs.(13) and (14),

$$|\varepsilon_e - \varepsilon_{hh}| \ll \frac{\hbar P}{m_0 a_0} \qquad (16)$$

In III-V materials $P_1$ and $P_2$ are in the interval $(1.1-1.5)\times 10^{-19}$ CGS and $a_0 \approx 0.55-0.65$ nm [25]. $P$ is smaller than $P_1$ and $P_2$ because the overlap integrals in Eq.(2) can be small. So a rough estimate gives $\hbar P/m_0 a_0 \approx 1$ eV.

**Discussion**

According to Eqs.(13) and (14) the width of the transition interval where $\alpha(\omega)$ grows from a negligible to a finite value and the width of the peak of $\operatorname{Re}\chi(\omega)$ as a function of $|\varepsilon_e - \varepsilon_{hh}|$ is controlled by $\hbar/\tau$. If $\tau = 10^{-11}$ s then $\hbar/\tau = 6\times 10^{-2}$ meV. The dependence of $\alpha/\alpha_{\max}$ and $\operatorname{Re}\chi/(\operatorname{Re}\chi)_{\max}$ on $|\varepsilon_e - \varepsilon_{hh}|-\hbar\omega$ near the threshold at this value of $\hbar/\tau$ is shown in Fig.2. Maximal absorption coefficient of the whole structure is $\alpha_{\max} = \alpha_{dw,\max} N$ where $N$ is the number of periods and minimal transmission coefficient above the absorption threshold is

$$T=(1-\alpha_{dw,max})^N \approx e^{-\alpha_{max}}.\qquad(17)$$

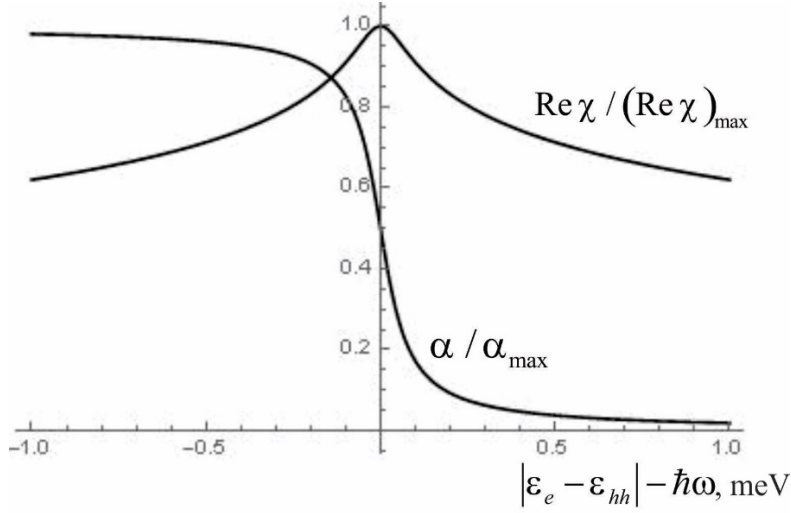

Fig.2. Dependence of $\alpha/\alpha_{max}$ and $\mathrm{Re}\,\chi/(\mathrm{Re}\,\chi)_{max}$ at $\hbar\omega=100\,\mathrm{meV}$ on $|\varepsilon_e-\varepsilon_{hh}|-\hbar\omega$ near the threshold for $\tau=10^{-11}$ s.

The minimal transmission coefficient defines the modulation depth, $m=(1-T)/(1+T)$ that does not depend on the design of the double well but only on the number of periods. The material and width of the layers of the double well control the absorption threshold and the magnitude of the field effect. That is if the average refractive index is 3.5 then to obtain modulation depth 90% (m=0.9) the structure has to have 450 periods. The change of the absorption coefficient from $0.1\alpha_{max}$ to $0.9\alpha_{max}$ takes place in interval of $|\varepsilon_e-\varepsilon_{hh}|\approx 0.35$ meV. If the distance between the middle of the wells is 5 nm then the change of $|\varepsilon_e-\varepsilon_{hh}|$ by this value requires the change of the electric field by about 700 V/cm.

Fig. 2 shows that $\mathrm{Re}\,\chi$ falls off rather slowly away from the peak because its dependence on the energy is logarithmical. So that $\mathrm{Re}\,\chi/(\mathrm{Re}\,\chi)_{max}=0.31$ at $||\varepsilon_e-\varepsilon_{hh}|-\hbar\omega|=10$ meV. But the value of $(\mathrm{Re}\,\chi)_{max}$ is quite large. For $\hbar\omega=100\,\mathrm{meV}$, $\tau=10^{-11}$ s, d=12 nm and n=3.5 Eq.(13) gives $(\mathrm{Re}\,\chi)_{max}=0.71$. This means that when $|\varepsilon_e-\varepsilon_{hh}|$ goes away from its resonance value by 10 meV the refractive index decreases from 4.6 to 3.87. If the distance between middle of the wells is 5 nm then change of $|\varepsilon_e-\varepsilon_{hh}|$ by 10 meV is reached by change of the electric field by about 20 kV/cm.

In summary we propose to use superlattices of double quantum wells separated by barriers for effective and high speed optical modulation and switching. The band gap of the double quantum wells can be controlled by electric field. In the vicinity of the optical threshold a small change of the electric field induces a sharp jolt of the absorption coefficient from a negligible to a finite value. The refractive index has its peak at the threshold and falls off significantly away from it. The dependence of the refraction index on the electric field can find its application in interference devices.

**Acknowledgement.**

This work was supported by National Science Foundation grant no. DMR-1809708, U.S. Army Research Office grant no. W911NF2010109.